\documentclass[english,pre,twocolumn]{revtex4}
\usepackage[normalem]{ulem}
\usepackage[T1]{fontenc}
\usepackage[latin9]{inputenc}
\usepackage{graphicx}
\usepackage{babel}
\usepackage{amssymb,amsmath}
\usepackage{subfigure, color}
\begin{document}

\global\long\def\pgr{\mathcal{P}_{\text{gr}}}
\global\long\def\pdb{\mathcal{P}_{\text{db}}}
\global\long\def\pov{\mathcal{P}_{\text{ov}}}
\global\long\def\pn{\mathcal{P}_{0}}
\global\long\def\df{d_{\text{f}}}
\global\long\def\DCmax{\Delta G_1}
\global\long\def\Rv{\mathcal{R}_v}
\global\long\def\ndr{n_{\text{dir}}}

\newcommand{\ie}{\emph{i.e.}}
\newcommand{\eg}{\emph{e.g.}}
\newcommand{\ER}{Erd\H{o}s-R\'{e}nyi}
\newcommand{\AP}{Achlioptas prcoesses}
\newcommand{\X}{\mathcal{O}}
\global\long\def\ploc{p_{\text{loc}}}

\title{
{
Genuine Non-Self-Averaging and Ultra-Slow Convergence in Gelation 
}
}

\author{Y. S. Cho$^1$, M. G. Mazza$^2$, B. Kahng$^1$, J. Nagler$^{3,2,1}$}

\affiliation{$^1$Center for Complex Systems Studies and CTP,Department of Physics and Astronomy, Seoul National University, Seoul 151-747, Korea}
\affiliation{$^2$Max Planck Institute for Dynamics and Self-Organization (MPI DS) G\"ottingen, Am Fa{\ss}berg 17, 37077 Germany}
\affiliation{$^3$Computational Physics, IfB, ETH Zurich, Wolfgang-Pauli-Strasse 27, 8093 Zurich, Switzerland}

\begin{abstract}
In irreversible aggregation processes droplets or polymers of microscopic size successively coalesce  until a large cluster of macroscopic scale forms. This gelation transition is widely believed to be self-averaging, meaning that 
the order parameter (the relative size of the largest connected cluster)
attains
well-defined values upon ensemble averaging with no sample-to-sample fluctuations in the thermodynamic limit. 
{
Here, we report on anomalous gelation transition types.
Depending on the growth rate of the largest clusters,
the gelation transition can show very diverse patterns as a function of the control parameter, which includes 
multiple stochastic
discontinuous transitions,
genuine non-self-averaging and ultra-slow convergence of the transition point. 
}
Our framework may be helpful in understanding and controlling gelation. 
\end{abstract}

\maketitle

\section{Introduction}

Irreversible aggregation phenomena are found in a great variety of physical and chemical systems such as polymerization reactions, 
antibody-antigen reactions, soot formation and gelling systems \cite{flory41,stockmayer43, book}. 
In the early 20th century Smoluchowski \cite{sm1916} studied such kinetic aggregation processes intensively and formulated a rate equation 
for the cluster densities in the mean-field approximation  as follows: 
\begin{equation}\label{SmEqu}
\frac{\text{d}n_k}{\text{d}t} = \sum_{i+j=k} K_{ij} n_i n_j- 2 n_k \sum_j K_{kj} n_j, 
\end{equation}
where $n_k$  is the density of $k$-size clusters and $K_{ij}$ is called the collision kernel 
that accounts for the adhesion of two clusters.
 The rate $K_{ij}n_i n_j$ is the probability that two clusters of sizes $i$ and $j$ merge
  per unit time and produces a cluster of size $k=i+j$. The negative term represents the case that a $k$-size cluster merges with any of the remaining clusters. The solution of the rate equation is still unknown for the great majority of collision kernels $K_{ij}$.
Linear polymerization 
 where two clusters are merged by the molecules at two reactive ends
 has been modeled by a constant kernel $K_{ij}$.
 When two clusters have a compact shape and merge, the kernel is given as $K_{ij}\sim (ij)^{1-1/d}$, 
 where $d$ is the spatial dimension. 
We consider the more general case of 
the power-law form $K_{ij} \sim (ij)^{\omega}$, which has attracted considerable attention since this form accounts for a great variety of aggregation processes. Examples include models with $0.5< \omega \le1$ that account for the effect of steric hindrance and intramolecular bonding \cite{leyvraz1982,leyvraz2003,ziff1982,ziff1983}.
Below the critical value $\omega_{\text{c}}=0.5$ aggregation based on Eq. (\ref{SmEqu}) exhibits a violation of mass conservation together with a lack of gelation in finite time, which has triggered a 
large body of work on the extensions (and corrections) of the Smoluchoswki's rate equation approach \cite{leod1962, book}.
In particular, it has been proven that
gelation for the (normalized) power law kernel 
${K}_{ij} =(ij)^{\omega}/(\sum_{s}s^{\omega}n_s)^2$ is continuous for $\omega > 0.5$ and
discontinuous for $\omega\leq 0.5$ \cite{cho2010, cho2011}. 

A constant kernel exponent $\omega$, however,  ignores a possible dependence of $\omega$ on 
the size (or surface) of the collision clusters \cite{ziff1983, herrmann1987}, for example, 
{
in the presence of 
effects highly
specific to the cluster sizes. 
Examples include cluster aggregation where rotation or gravitation leads to mass segregation
as discussed in Appendix B.
Here we demonstrate that if the collision 
rates of the {\em largest} cluster (or the largest clusters) are controlled (by intrinsic or extrinsic effects), 
gelation can exhibit anomalous critical and supercritical behaviors \cite{nagler2015, jia2016}. 
}

\begin{figure}[t]
\includegraphics[width=0.8\linewidth]{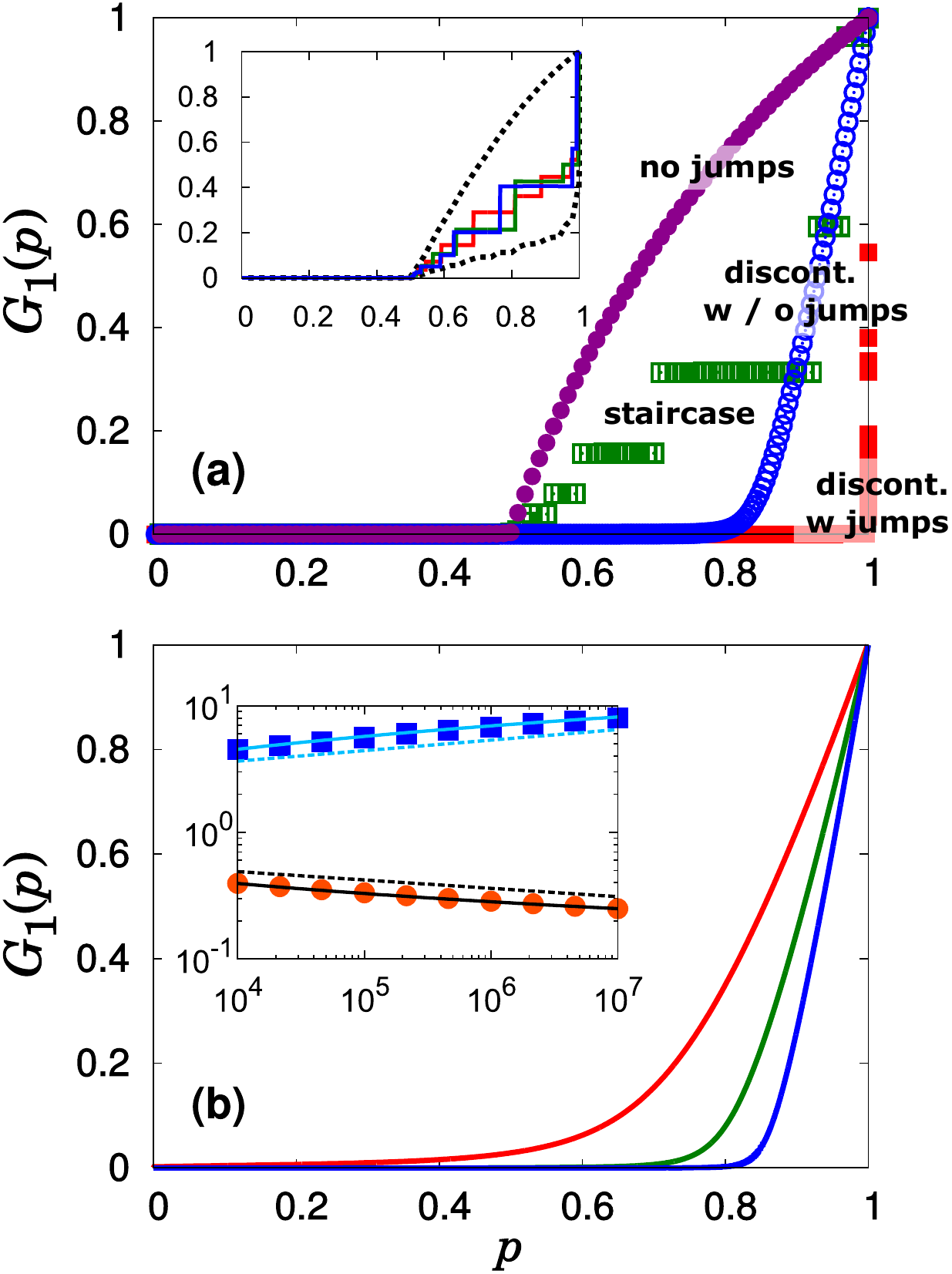}
\caption{(Color online)\label{fig:sketch}
{\bf Single realizations exemplify various gelation types.}
(a)
The relative size of the largest component $G_1(p)$ vs $p$ for single configurations
with $N = 2 \times 10^6$. 
We find four gelation types I-IV.  In the case of 
 type I ($\bullet$, purple), $G_1(p)$ for $(\alpha,\beta)=(1,1)$ 
shows continuous transition at $p_{c}<1$ and exhibits 
no jumps. 
For type II ($\square$, green), $G_1(p)$ at $(1,0)$ follows a staircase. 
The steps of the staircase are stochastic,  even for $N\rightarrow\infty$. 
In the inset, we show single realizations at $(1,0)$ from different configurations and {envelopes} of those staircase patterns 
(dotted curves) for $N=10^7$. The {envelopes} are given by the minimum and maximum, respectively, of the realizations
as a function of $p$.
For type III ($\blacksquare$, red), 
$G_1(p)$ at (0,0) shows discontinuous transition, induced by single step gaps, at $p_{c} = 1$.
For type IV ($\circ$, blue), $G_1(p)$ at (0,1) exhibits a continuous transition at the transition point $p_c$, 
which moves ultra-slowly to $p_c=1$ for $N \to \infty$. 
(b) Ultra-slow convergence:
We show $G_1(p)$ vs $p$ at $(0,1)$ for different system sizes 
$N = 10^3$, $10^5$, and $10^7$. Inset:
Plot of $1-p_{c}$ ($\bullet$) and $dG_1/dp|_{\text{max}}$ vs $N$ $(\blacksquare)$ are shown.
Dashed lines follow power law with exponents $-0.07$ and $0.08$.
Solid lines are ${\log(x)}^{-0.82}$ and ${\log(x)}^{1.05}$ for comparison.}
\end{figure}


\section{Model}

We study the rate equation (\ref{SmEqu}) for a composite collision kernel that differs only
 from the normalized power-law kernel in the growth rate of the largest cluster, which is given as 
follows: $K_{ij}=k_ik_j$, where $k_i=i^{\omega}/{\mathcal{N}}$ with the normalization constant $\mathcal{N}=\sum_{s=1}^{S_1-1}s^{\alpha}n_s+S_1^{\beta} n_{S_1}$,
\begin{eqnarray}\label{eq:compositekernel}
\omega = 
\begin{cases} 
\alpha  &\mbox{if } i \ne S_1, \\ 
\beta   & \mbox{otherwise}
\end{cases} 
\end{eqnarray}
where $n_s$ is the density of 
 clusters of size $s$, and $S_1$ is the size of the largest cluster in the system.

These composite kernels introduce 
a separation of time scales 
occurring in a number of simple physical systems such as 
diffusion-limited aggregation under gravity and
cluster growth 
in a linear shear profile.
The cause for segregation in those systems is a substantially different 
growth rate for very large aggregates, compared to smaller clusters.
For example large clusters can move to the bottom of a vessel or can be driven towards regions
where the local cluster size distribution differs from those of other places.
This behavior can be most pronounced when an infinite cluster is about to emerge or has emerged
 (see Appendix B). 

We perform kinetic Monte Carlo simulations in the following way.
Starting with $N$ monomers of size one, 
  each time step 
 two 
  clusters of sizes $i$ and $j$ are randomly selected  
with the weight $K_{ij}$ given by Eq.~({\ref{eq:compositekernel}}) and are merged. 
Next, the control parameter $p$ (the normalized time) is increased by $\Delta p=1/N$, which ensures $p \leq 1$.
Gelation is 
determined
by studying the 
order parameter $G_1(p)\equiv S_1(p)/N$, the relative size of the
largest cluster, as a function 
%
of $p$ which characterizes the gelation transition from microscopic connectivity (the sol) to macroscopic connectedness (the gel) in the thermodynamic limit $N\rightarrow\infty$. 

\begin{figure}[t]
\includegraphics[width=0.6\linewidth]{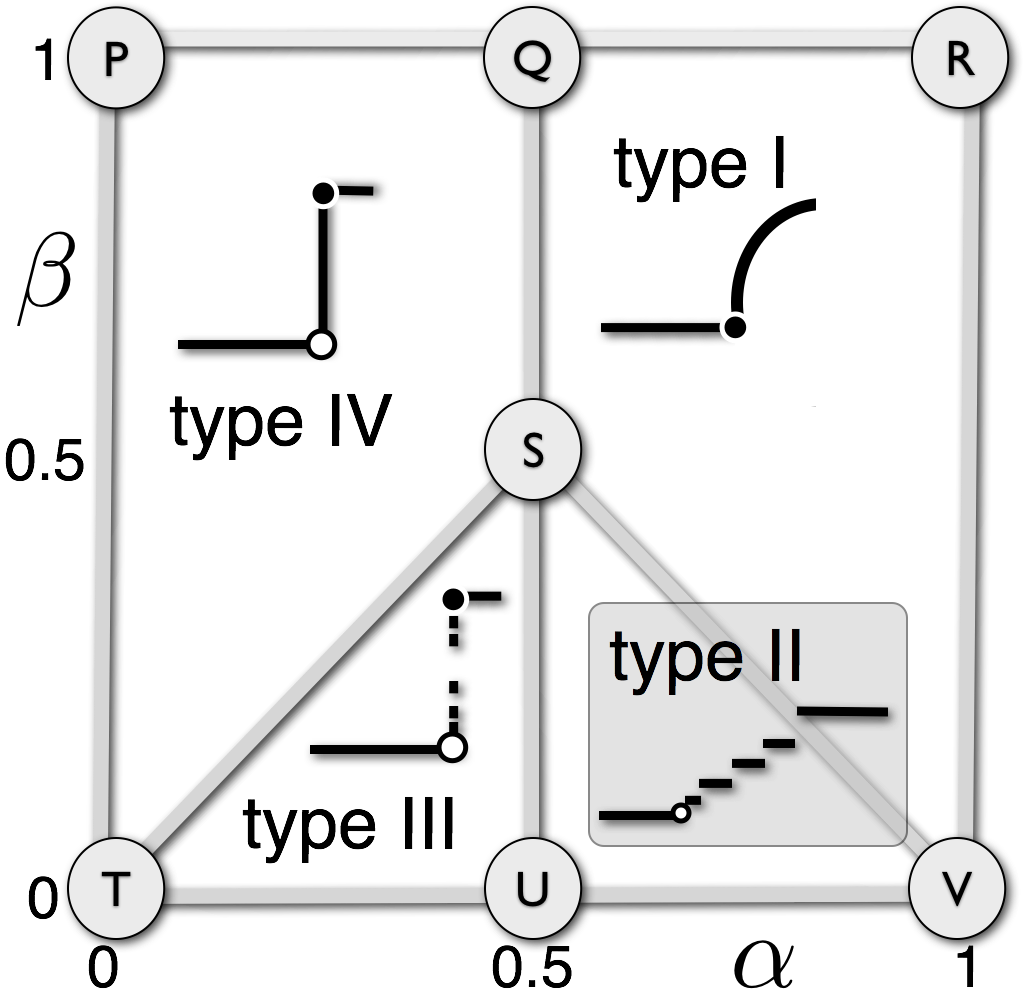}
\caption{(Color online)
{\bf Phase diagram of the gelation types in the plane $(\alpha, \beta)$.} 
The transition point is classified into two cases by location, $p_c <1$ for $\alpha > 0.5$ and $p_c=1$ for $\alpha \leq 0.5$.
{
Genuine non-self-averaging behavior appears in the region of type II: 
not only the ensemble of realizations are stochastic in the supercritical regime (shaded triangle) but also
singe realizations become fully stochastic in the thermodynamic limit (see text).
}
The ultra-slow converging behavior of the transition point appears in the region of type IV. 
}\label{fig:phasediagram}
\end{figure}

\section{Results}

Single realizations of the evolution of the largest cluster illustrate the occurrence of four gelation types (I-IV) for different combinations of the exponents $\alpha$ and $\beta$ (Fig.~\ref{fig:sketch}). The phase diagram of those different types in the plane of ($\alpha,\beta$) is shown in Fig.\ \ref{fig:phasediagram}.

\begin{figure}[t]
\includegraphics[width=0.8\linewidth]{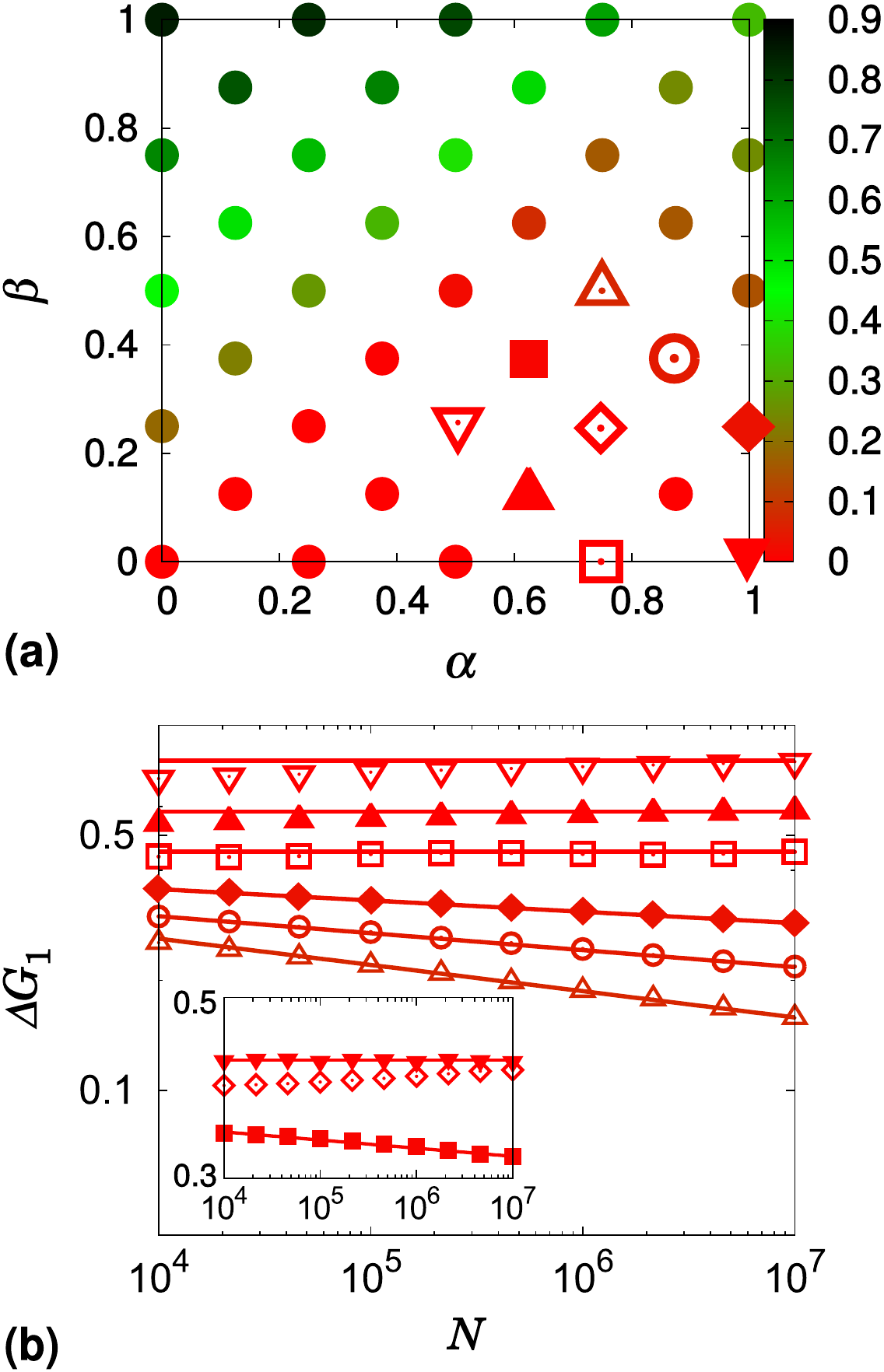}
\caption{(Color online) \label{fig:scaling}
{\bf Largest gap statistics.}
(a) Plot of the exponent $\delta$ of $\Delta G_1 \sim N^{-\delta}$ in the plane of $(\alpha, \beta)$. 
Numerics supports that $\delta > 0$ for types I and IV and $\delta = 0$ for types II and III. 
(b)
For 
characterization of the boundary between type I and II 
we plot 
$\Delta G_1$ vs $N$ for the three points (0.75,0.5) $(\triangle)$, (0.875,0.375) $(\circ)$, (1,0.25) $(\blacklozenge)$ in 
the region of type I. They decay with increasing $N$ with the slopes $-0.07$, $-0.05$ and $-0.03$, respectively. We also checked the cases (0.5,0.25) $(\triangledown)$, (0.625,0.125) $(\blacktriangle)$ and (0.75,0) $(\square)$ in the region of type II. $\Delta G_1$ seems to be independent of $N$. 
For visibility 
the data for $(\blacktriangle, \triangledown)$ are shifted upward. 
Inset: 
Plots of {$\Delta G_1(N)$} vs $N$ for the data points (0.625,0.375) $(\blacksquare)$, (0.75, 0.25) ($\Diamond$), and (1.0, 0) ($\blacktriangledown$) on the boundary SV suggests a marginal behavior, that is, $\Delta G_1(N)$ displays an ultra-weak dependence on $N$.
Slopes of the guidelines are -0.01 and 0.
}
\end{figure}

\paragraph{Type I:} We characterize type I as globally continuous gelation, implying  a continuous  transition of $G_1(p)$ at the critical point $p_c<1$ (Fig.~\ref{fig:sketch}) and a vanishing maximal one-step gap in $G_1$, $\Delta G_1:=\max_p(G_1(p+1/N)-G_1(p))\rightarrow 0$ (Fig.\ \ref{fig:scaling}). 
This method has been helpful to
distinguish between continuous and discontinuous 
 percolation models \cite{achlioptas, nagler2011, ziff2009, raissa2010, riordan2011,naglerPRE1,naglerPRE2, nagler2016}.
Type I is found in the phase diagram in the domain QRVS given by $\alpha > 0.5$ and $\alpha+\beta>1$ (Fig.\ \ref{fig:phasediagram}).

\begin{figure}[t]
\includegraphics[width=0.8\linewidth]{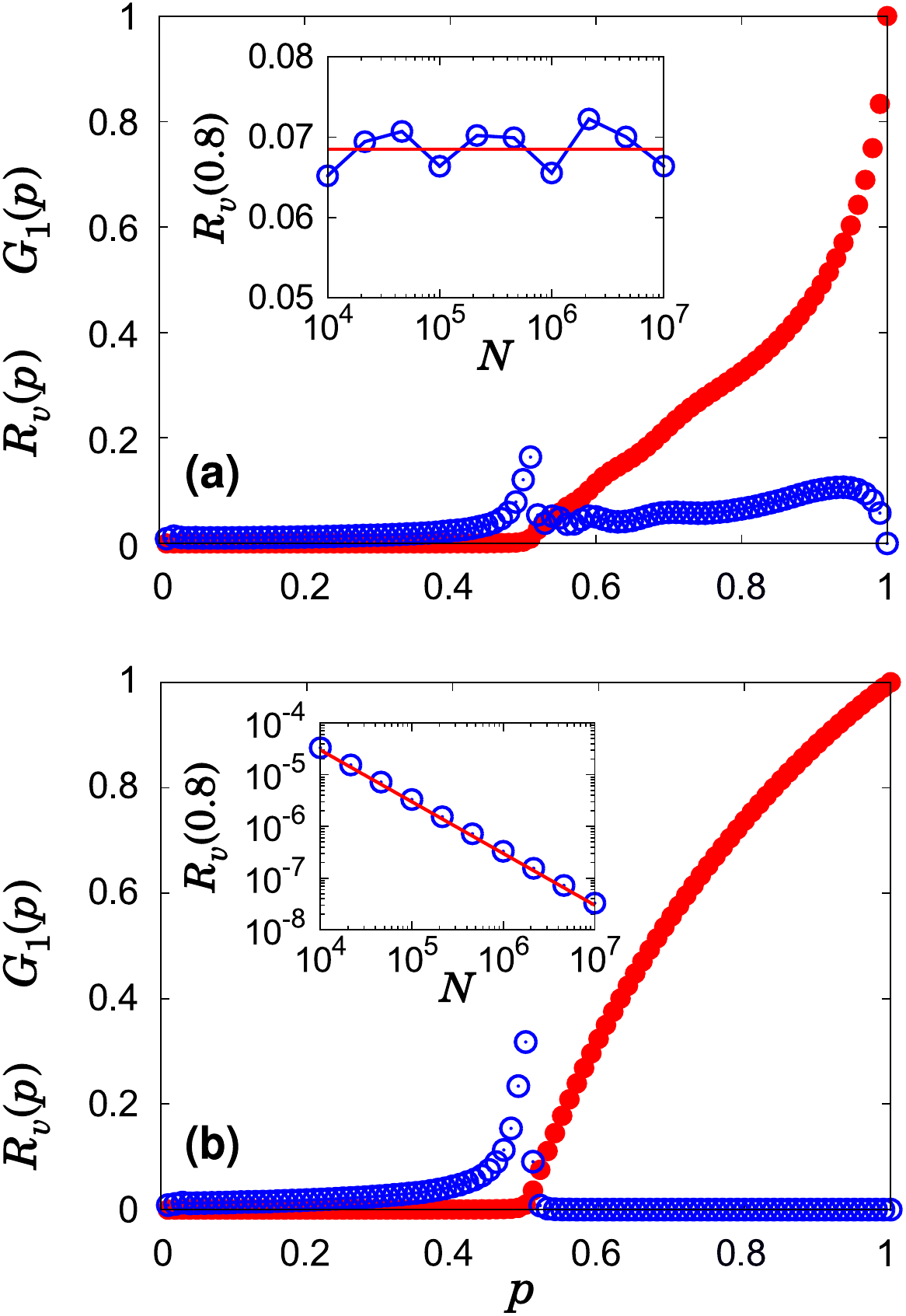}
\caption{(Color online)\label{fig:Rv}
{\bf Non-Self-averaging for type II vs self-averaging for type I.}
(a) $G_1(p)$($\bullet$, red) and the relative variance $R_{\text{v}}(p)$($\circ$, blue) vs $p$ for $(\alpha, \beta)$=(1,0) in type II.
$R_{\text{v}}(p)>0$ when $p>p_{c}$, which suggests non-self-averaging.  
Data are obtained from $N=10^6$ averaged over $2\times 10^4$ realizations.
Inset: To test for non-self-averaging, we plot  $R_{\text{v}}$ vs $N$ at $p=0.8$. 
$R_{\text{v}}$ oscillates near $0.069$, which suggests that $R_{\text{v}}$ does not shrink to zero 
in the thermodynamic limit. (b) $G_1(p)$ ($\bullet$, red) and $R_{\text{v}}(p)$ ($\circ$, blue) vs $p$ for $(\alpha, \beta)=(1,1)$ in type I. 
$R_{\text{v}}(p)=0$ when $p>p_{c}=0.5$, which characterizes self-averaging.
Inset: To test the self-averaging, we plot $R_{\text{v}}$ vs $N$ at $p=0.8$.
$R_{\text{v}}$ decays to zero as $\sim 1/N$.}
\end{figure}

{
\paragraph{Type II:} 
For the region satisfying $\alpha+\beta < 1$ and $\alpha > 0.5$, 
the order parameter in a single realization follows a staircase beginning at the critical point $p_c <1$. 
Specifically, there exist multiple, finite one-step gaps in $G_1$ for $N\rightarrow\infty$. 
This pattern has been observed in percolation \cite{riordan2012, nagler2012, nagler2013}. 
The plateaus are caused by the stagnation of the growth of the giant component. During 
 finite intervals, other clusters can grow to $\it{O}(N)$ and can aggregate with the largest component and cause multiple finite jumps of the order parameter \cite{nagler2012}. Intriguingly, the positions of the staircase steps are randomly distributed, even in the thermodynamic limit.
To see this, we study the relative variance of the order parameter
$R_{\text{v}}(p)=\frac{\langle G_1^2 (p) \rangle - \langle G_1(p)\rangle^2}{\langle G_1(p)\rangle^2}$ which does not scale away for $N\rightarrow\infty$,
for $p>p_c$ (Fig.\ \ref{fig:Rv}(a)).
This means that the order parameter in the type II phase transition does not converge to a function $G_1(p)$ for $p>p_c$.
This behavior stands in contrast to usual self-averaging gelation processes (Fig.\ \ref{fig:Rv}(b)) 
but has reported earlier in models of random network percolation \cite{riordan2012, nagler2013}.

Type II, however, differs qualitatively from previous reported stochastic staircases
 in random network percolation \cite{riordan2012, nagler2013} where single realizations of $G_1$ necessarily
 jump instantaneously to the upper {envelope} when touching the lower one \cite{nagler2015}.
 This rather unphysical behavior is a consequence of
  the strict impossibility for $G_1$ to grow unless the second largest cluster has exactly the same size as $G_1$,
  a built-in mechanism of the models studied in Refs.\ \cite{riordan2012, nagler2012, nagler2013}.
  
By contrast, the staircase of type II is {\em fully} stochastic in the supercritical regime, for $p>p_c$:
not only the ensemble but also single realizations are stochastic (even for $N\rightarrow \infty$).
This behavior (referred here to as {\em genuine} non-self-averaging) 
is the behavior typically occurring in spin glasses.
}

\paragraph{Type III:}
We characterize type III by a single  discontinuous transition at the end of the process, $p_c=1$ \cite{cho2010,araujo,sca, deterministic} 
together with the occurrence of a finite gap induced by a single merger, $\Delta G_1\rightarrow \text{const.}>0$ for $N\rightarrow\infty$ \cite{nagler2011}. 
We find Type III to occur in the region $\beta < \alpha \leq 0.5$ of the phase diagram (Fig.\ \ref{fig:phasediagram}). The special case $\alpha=\beta<0.5$
was reported earlier  \cite{cho2010}.

\paragraph{Type IV:}  
The type IV transition occurs in the domain PQST $(\alpha \leq 0.5,\alpha<\beta)$ of the phase diagram  (Fig.\ \ref{fig:phasediagram}). As shown in Fig.~\ref{fig:sketch}, $G_1(p)$ seems to exhibit a continuous transition. However, the finite size transition point $p_c(N)$ approaches $p_c=1$ as $N$ is increased. 
Specifically, 
for $(\alpha,\beta)=(0,1)$
the approaching rate is ultra-slow characterized by { $1-p_c(N) \sim N^{-0.07}$}, 
together with an ultra-slow increase of the maximum slope { $dG_1/dp|_{\text{max}} \sim N^{0.08}$}. This behavior is shown in the inset of Fig.~\ref{fig:sketch}(b).
Accordingly, $G_1(p)$ sharply increases at $p_c=1$ in the limit $N\to \infty$. Due to $p_c=1$ this transition 
is discontinuous but still lacks a one-step gap.   For fixed $N$,
the point {P} exhibits the smallest single step gap size in the domain PQST (Fig.\ \ref{fig:phasediagram}). 

To further substantiate our claims 
we have performed an extensive scaling analysis 
 of the size of the largest gap \cite{nagler2011}
\begin{equation}\label{eq:delta}
\Delta G_1 \sim N^{-\delta}.
\end{equation}
Fig.~\ref{fig:scaling}(a) shows the $(\alpha,\beta)$ plane where the color codes for $\delta$, based on Eq.\ (\ref{eq:delta}). The numerics suggests that the largest gap scales away for $N\rightarrow\infty$ for the 
domains of types I and IV. Our numerics also supports position and extent of discontinuous transitions of 
type II or III ($\delta=0$) as shown in Fig.\ 2.

{\it Discussion.$-$}
{
Gelation can show anomalous behaviors 
when 
two (or more)  coalescence time scales compete. 
Examples include 
systems  where 
a force opposes diffusion in one spatial direction,
as for aggregation processes  affected by gravity, 
or in rotating (planetary ring) systems (see {Appendix B}, and Ref.\ \cite{saturn2015} for a recent study 
on Saturn's rings coalescence dynamics).

Ultra-slow convergence towards $p_c=1$ with no genuine single step gaps (type IV), occurring in an extended parameter regime, 
represents an anomalous phase transition type of discontinuous gelation \cite{waage, nagler2015}. 
This behavior can be related to aggregation under linear shear and other collision kernels (see {Appendix B}).

Genuine non-self-averaging behaviors are qualitatively different from non-self-averaging found in previous models where
a finite number of genuine jumps of the order parameter imply the non-self-averaging \cite{bennaim2005},
or from anomalous supercriticality previously reported in random network percolation.
We here showed when and how 
multiple discontinuous transitions and stochastic staircases in gelation (type II) arise
from a truly stochastic dynamics in the supercritical regime (and not merely due to frozen random events
at exactly $p_c$, determining the 'phase' of the staircase as in recently introduced models \cite{riordan2012, nagler2013, nagler2015}).

Controlling the largest $m$ clusters (instead of $m=1$) leads to the same phenomenology (in particular, types II \& IV,
 see {Appendix A}). This demonstrates the robustness of our results.
 
Non-self-averaging necessarily implies large sample-to-sample fluctuations 
during the gel formation and avoidance is therefore crucial for controlling gelation.
Anomalous supercritical behaviors are expected in percolation and cluster aggregation with
a separation of the reaction time scales, in particular due to mass segregation.
Future work must establish how 
large sample to sample fluctuations induced by this effect occur in experiments.

In contrast to previous work where 
diverse phenomena in cluster merging processes have been observed and explained with different methods,
the present work is an attempt to unify anomalous phenomena in gelation.

}

\section*{ACKNOWLEDGMENTS}

This work was supported by the NRF-ERC exchange
program (grant no.\ 2010-0015066), National Creative Research
Initiative (grant no.\ 2014-069005), and SNU R\&D
grant (BK) and the Global Frontier Program (YSC).
JN acknowledges financial support from the ETH Risk Center (SP RC 08-15)
and thanks C. Comosum and E. Trigona for advice.

\section*{APPENDIX}

\subsection{Observation of four types of phase transitions in controlling the largest clusters}

To study the robustness of our model, 
we consider a generalized model to control the growth of the $m$ largest clusters in the entire system. 
Here, we modify the collision kernel of Eq.~(2) as 

\begin{eqnarray}\label{eq:compositekernel2}
\omega = 
\begin{cases} 
\alpha  &\mbox{if } i \notin R_m,\\ 
\beta   & \mbox{if } i \in R_m,
\end{cases} 
\end{eqnarray}
where $R_m$ is the set of $m$ largest clusters in a  given configuration. If there are multiple clusters of size $S_m$, 
we randomly select one among them,
 where $S_m$ is the size of $m$-th largest cluster. We remark that $m=1$ in this modified model is different from the original model as well because there can be numerous clusters of size $S_1$.
 
\begin{figure*}[t!]
\includegraphics[width=0.95\linewidth]{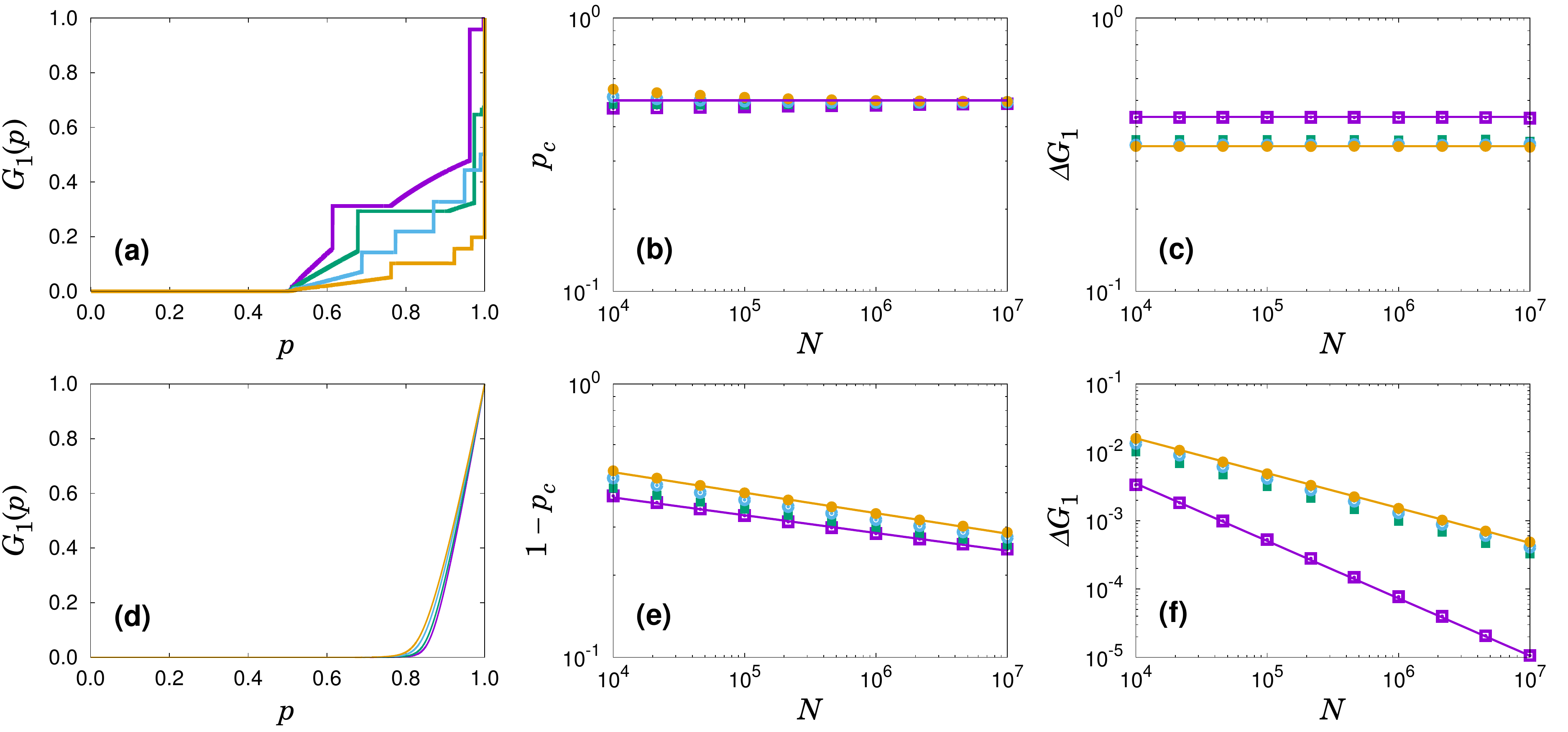}
\caption{
{\bf Controlling the largest $m$ components leads to 
type II (a-c)  and type IV (d-f).}
We use $m = 1, 2, 5$ and $10$ (See the text for definition.) for ($\alpha, \beta$)=(1,0) (a-c) and (0,1) (d-f). (a) $G_1(p)$ vs $p$ with $N = 10^7$. As $m$ increases, $G_1(p)$ increases more
drastically near $p=1$.  Irrespective of $m$, $G_1(p)$ 
behaves as a stochastic staircase (type II). 
(b) $p_c(N)$ vs $N$ for $m = 1, 2, 5$ and $10$ from the below. $p_c(N)$ decreases to some finite value $p_c(\infty)<1$ within the simulation range. Here, $p_c(N)$ is taken as ${\textrm {argmin}}_p(S_1 \geq N^{1/2})$ same with $p_c(N)$ used in the {\textcolor{black}{Fig.~\ref{fig:sketch}}}. (c) The maximal jump size of $G_1$ vs $N$ for $m=1, 2, 5$ and $10$ from the above. $\Delta G_1$ is independent of $N$ within the simulation range. (d) $G_1(p)$ vs $p$ with $N = 10^7$ for $m = 1, 2, 5$ and $10$ from the right. (e) $1-p_c(N)$ vs $N$. The slopes of guidelines are $-0.075$ and $-0.065$ from the above. (f) $\Delta G_1$ vs $N$. The slopes of guidelines are $-0.84$ and $-0.51$ from the below.
}\label{figS:multilargest}
\end{figure*}

We are interested in whether this modified model shows all four transition types. Irrespective of $m$, it is obvious that this model is equivalent to the original model when $\alpha = \beta$. Thus, it is already shown that this model shows type I transition when $(\alpha, \beta) = (1, 1)$ and type III transition when $(0, 0)$. Next, 
we show the  
test of the parameter choice $(1, 0)$ and $(0, 1)$ for types II and IV in Fig.~\ref{figS:multilargest}. 
Again, we use $p_c(\infty)<1$ and $\Delta G_1(\infty) > 0$ as the  criterion to identify type II transitions and $p_c(\infty)=1$ and $\Delta G_1(\infty) = 0$
as the criterion to identify type IV transitions. It is confirmed numerically that type II (type IV) transition is observed when $(1, 0)$ $((0, 1))$ for $m = 1, 2, 5$ and $10$ as shown in Fig.~\ref{figS:multilargest}.  We expect that this result 
generalizes to arbitrary values of $m$.

\subsection{Physical systems exhibiting anomalous transition types}
 
\subsubsection{Diffusion-limited cluster aggregation}

Diffusion-limited cluster aggregation (DLA)
was originally suggested to model the formation of fractal structure of aggregated particles~\cite{dlca1, dlca2} and 
allowed to study extensively 
dynamic properties such as the
cluster size distribution. 
Experimental realizations were achieved 
by aggregating  silica microspheres floating on salty water~\cite{dlca_exp}.
Diffusion-limited cluster aggregation has been  studied in the context of percolation transition for the first time in~\cite{cho2011} 
and the authors found that the model shows a discontinuous percolation transition.
Here, we study this model in more detail from the perspective of percolation and clarify which type of transition is observed in this model. 

\begin{figure}[t!]
\includegraphics[width=1.0\linewidth]{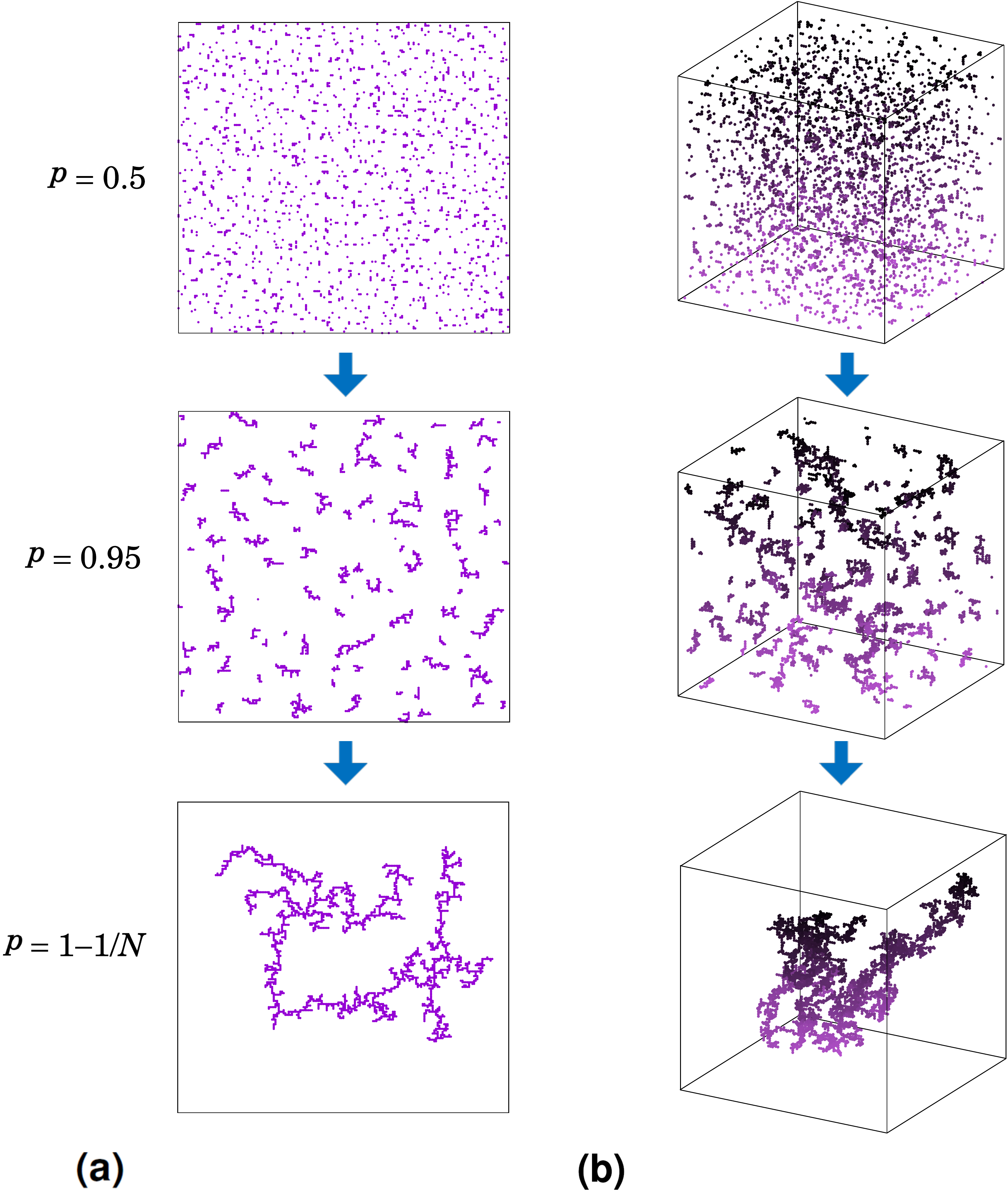}
\caption{
{\bf Growth of clusters in diffusion-limited cluster aggregation.}
This figure shows the snapshot of clusters following diffusion-limited cluster aggregation in two (a) and three (b) dimensions 
with $p=0.5$, $0.95$ and $1-1/N$ from {\textcolor{black}{the top to bottom in  each column}}. $L=2\times10^2$ and $N=2\times10^3$ are used for (a) and
$L=10^2$ and $N=4\times10^3$ are used for (b). The color of element in (b) varies continuously from purple to black as $z$ coordinate increases.}\label{figS:dlcasnapshot}
\end{figure}

The model describes aggregation between mobile clusters. 
Clusters move following a Brownian motion in $d$ dimensional space, and aggregate
with each other when they are adjacent. By the property of the Brownian motion, the mean velocity of the clusters follow $v_s \sim \sqrt{dkT/s}$, where $s$ is size of the clusters and $T$ is temperature. To simulate this model, we use the following method. At $p=0$, we distribute $N$ isolated nodes in $d$ dimensional square lattices of length $L$. We remark that no pairs of nodes are adjacent and all nodes are isolated clusters at the beginning. Then, at each time step, one cluster of size $s$ is selected with probability proportional to $1/\sqrt{s}$ and moves one unit to one of $2d$ directions randomly. Then, two different clusters can be placed at the nearest neighbor positions and merge to form a larger cluster. If two clusters merge, $p$ is increased by $p\rightarrow p+1/N$. The order parameter $G_1(p)$ is the size of the largest cluster divided by $N$. To estimate the thermodynamic limit of this system,
 we increase the system size $N$ for fixed density of particles $\rho = N/L^d$.
In Fig.~\ref{figS:dlcasnapshot}, the snapshots of clusters 
for $d=2$ and $d=3$
 are shown. We find that the clusters have fractal structure. It is known that the fractal dimensions of the clusters are $d_f \approx 1.4$ for two dimensions and $d_f \approx 1.8$ for three dimensions.

\begin{figure*}[t!]
\includegraphics[width=0.95\linewidth]{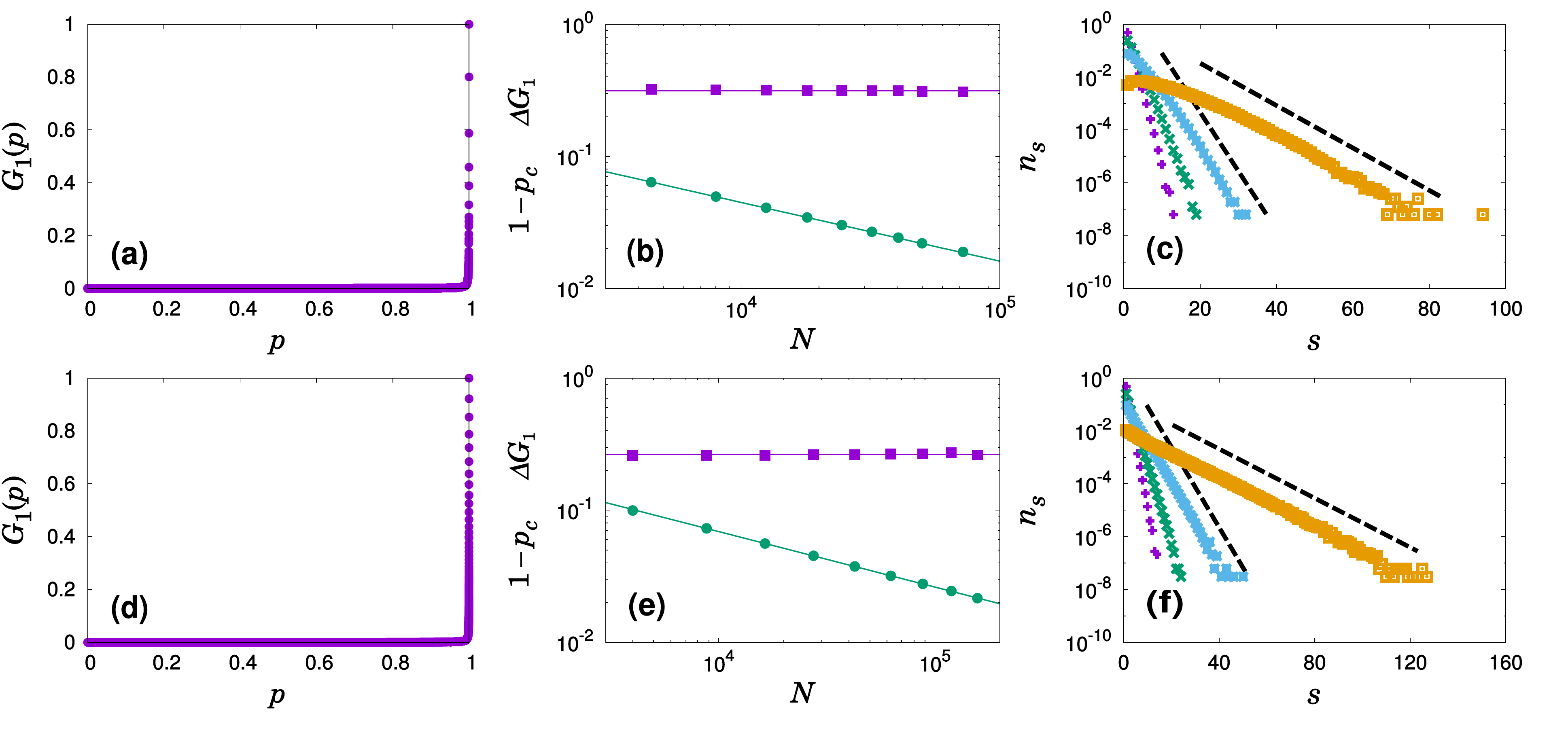}
\caption{
{\bf Type III transition in diffusion-limited cluster aggregation.}
(a) $G_1(p)$ vs. $p$ for $N=7.2\times10^4$ and $L = 1.2\times10^3$ in two dimensions. $G_1(p)$ increases drastically up to unity near $p=1$.
(b) $1-p_c(N) (\bullet) $ and $\Delta G_1 (\blacksquare)$ vs. $N$ for $\rho = 0.05$. $\Delta G_1$ is independent of $N$ and $1-p_c(N)$ 
decreases following a power law within the simulation range, which implies that type III transition occurs in the diffusion-limited cluster aggregation in two dimensions.
Here, $p_c(N)$ is taken as ${\textrm {argmin}}_p(S_1 \geq N^{1/2})$ same with $p_c(N)$ used in the {\textcolor{black}{Fig.~\ref{fig:sketch}}}. The slopes of guidelines are
$-0.44$ and $0$ from the below. (c) $n_s$ vs. $s$ for $p=0.3, 0.5, 0.7$ and $0.9$ from the left in the semi-log scale.  $N=8 \times 10^3$ and $L=4 \times 10^2$ are used. We can find that $n_s$ decreases
exponentially in large cluster region irrespective of $p$.
(d) $G_1(p)$ vs. $p$ for $N=1.08\times10^5$ and $L = 3\times10^2$ in three dimensions. $G_1(p)$ increases drastically up to unity near $p=1$.
(e) $1-p_c(N) (\bullet)$ and $\Delta G_1(\blacksquare)$ vs. $N$ for $\rho = 0.004$. $\Delta G_1$ is independent of $N$ and $1-p_c(N)$ decreases following power law within the simulation range, which implies that type III transition occurs in the diffusion-limited cluster aggregation in three dimensions.
The slopes of guidelines are $-0.42$ and $0$ from the below.
(f) $n_s$ vs. $s$ for $p=0.3,0.5,0.7$ and $0.9$ from the left in the semi-log scale. $N=2^{14}$ and $L=1.6\times10^2$ are used. We can find that $n_s$ decreases
exponentially in large cluster region irrespective of $p$.
}\label{figS:dlcatypeIII}
\end{figure*}

Now, we determine the type of transition in this process. As shown in Fig.~\ref{figS:dlcatypeIII}(a) and (d), $G_1(p)$ increases drastically at $p \approx 1$,
which means that a type III or type IV transition is expected in this model. To specify the transition type, we measure $1-p_c(N)$ and $\Delta G_1$ vs. $N$
as shown in Fig.~\ref{figS:dlcatypeIII}(b) and (e). Within the simulation range, $1-p_c(N)$ decreases to zero following a power law as $N$ increases, while $\Delta G_1$ is independent of $N$, which indicates that this model shows type III transition in both two and three dimensions.

To analyze this result within our theoretical framework, we study the behavior of the collision kernel in this model. It is known that the cluster aggregation process of this model may be described via an asymmetric Smoluchowski equation

\begin{equation}
\frac{\text{d}n_s}{\text{d}p}=\sum_{i+j=s}k_ik'_jn_in_j - n_sk_s - n_sk'_s,
\end{equation}
where $k_i \sim i^{1-1/d_f}$ and $k'_j \sim j^{1-1/d_f - 0.5}$~\cite{cho2011}. 
This is derived from the fact 
 that the effective surface area of cluster of size $i$ scales as 
 $i^{1-1/d_f}$.
When two clusters aggregate, one cluster is mobile and the other cluster is immobile. Thus, the collision kernel for aggregation of
clusters of size $i$ and $j$ may be written as the product of $k_i$ and $k'_j$, where $k_i$ is the collision kernel for immobile cluster and $k'_j$ is the collision kernel for mobile cluster. This behavior was checked numerically in~\cite{cho2011}.
We can obtain $k_i \sim i^{0.29}$, $k'_j \sim j^{-0.21}$ for two dimensions and $k_i \sim i^{0.45}$, $k'_j \sim j^{-0.06}$ for three dimensions by using known $d_f$ values. 
To relate these collision kernels to the $\alpha = \beta$ case of the collision kernel {\textcolor{black}{of Eq.~(\ref{eq:compositekernel})}}, we investigate $n_s$ of the diffusion-limited cluster aggregation process as shown in Fig.~\ref{figS:dlcatypeIII}(c) and (f). We find that $n_s$ decreases exponentially irrespectively of $p$, which means that
the cluster size distribution is not heterogeneous. This may be due to the fact 
that the exponents of both mobile and immobile collision kernels are smaller than $0.5$. 
Then, we use the approximation $k_i k'_j \sim i^{1-1/d_f}j^{1-1/d_f-0.5} \approx i^{1-1/d_f-0.25}j^{1-1/d_f-0.25}$ which is valid when $i \approx j$, 
because the cluster size distribution is not heterogeneous during the process. If we use this approximation, the dynamics of diffusion-limited cluster aggregation can be related to the collision kernels {\textcolor{black}{of Eq.~(\ref{eq:compositekernel})}} as $(\alpha, \beta)=(0.04, 0.04)$ for two dimensions and $(\alpha, \beta)=(0.20, 0.20)$ for three dimensions, where type III transitions
are observed. $n_s=(1-p)^2p^{s-1}$ for $(\alpha, \beta)=(0,0)$ was analytically obtained in~\cite{cho2010} and the exponentially decreasing behavior of $n_s$ was numerically checked 
{\textcolor{black}{for $(\alpha,\beta)=(0.04,0.04)$ and $(\alpha,\beta)=(0.20,0.20)$}}, which supports this analysis.

\begin{figure*}[t!]
\includegraphics[width=0.95\linewidth]{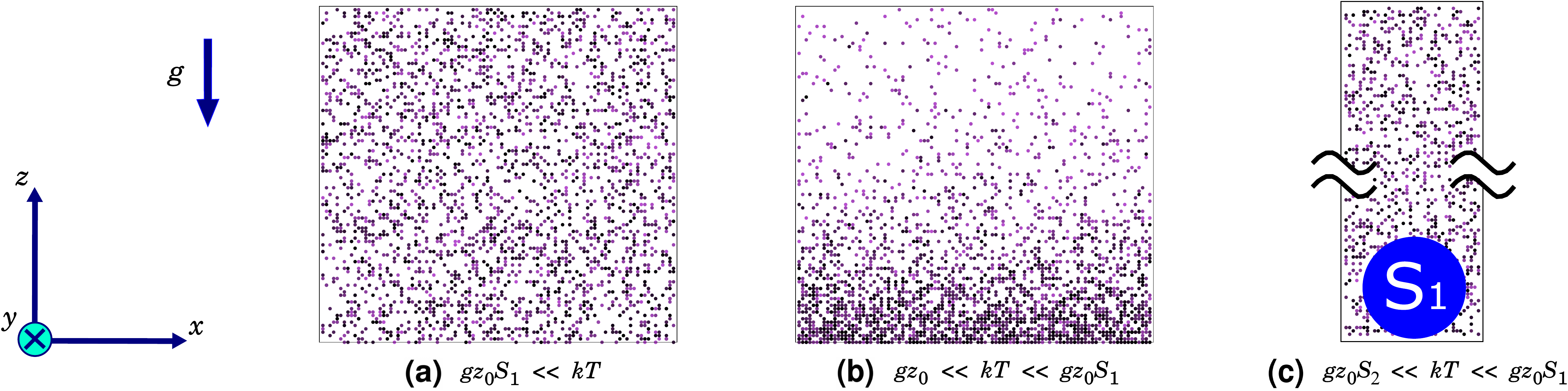}
\caption{
{\bf Schematic for diffusion-limited cluster aggregation in gravitational field.}
Finite clusters are represented by points for visualization.
Color of each cluster changes continuously from purple to black as its mass increases.
(a) Finite clusters are almost uniformly distributed irrespective of their sizes when $gz_0S_1 \ll kT$. (b) When $gz_0 \ll kT \ll gz_0S_1$, finite clusters of size $s$ for $gz_0s \ll kT$ are almost uniformly distributed but finite clusters of size $s$ for $kT \ll gz_0s$ are more densely populated as closed to the bottom $z=0$. 
(c) When there are finite clusters and one giant cluster of size $S_1$ (blue circle) in the condition $gz_0S_2 \ll kT \ll gz_0S_1$, the giant cluster moves randomly on the bottom $z=0$ and finite clusters are almost uniformly distributed.
}\label{figS:schematic_boltzmann}
\end{figure*}

\subsubsection{Diffusion-limited cluster aggregation in gravitational field}

Here we consider clusters in a vessel at temperature $T$ in the presence of 
a constant gravitational force pointing in the negative $z$-direction. Then, the density of clusters of size $s$
at position $z$ which is denoted by $n_{zs}$ follows the Boltzmann distribution,
\begin{equation}
n_{zs} \sim e^{-sgz/kT}
\end{equation}
as shown in the Fig.~\ref{figS:schematic_boltzmann}(a) and (b), where $kT$ is the thermal energy.
If the height of the vessel is $z_0$, normalization $\int^{z_0}_{z=0} n_{zs}dz =z_0n_s$ leads to 
\begin{equation}
n_{zs}=\frac{z_0n_s e^{-sgz/kT}}{\frac{kT}{sg}(1-{\textrm {exp}}(-sgz_0/kT))}.
\end{equation}
We assume that a collision rate between clusters of sizes $i$ and $j$ in a local region at position $z$ has the form $n_{zi}n_{zj}(ij)^{\alpha}$, because the distribution of clusters would be uniform and thus follows diffusion limited cluster aggregation locally. 
Then a total collision rate between clusters of sizes $i$ and $j$ is given as
\begin{equation}
K_{ij}n_in_j \sim (ij)^{\alpha}\int^{z_0}_{z=0}n_{zi}n_{zj}dz.
\end{equation}
When $gz_0S_1 \ll kT$, one can show easily $K_{ij} \sim z_0(ij)^{\alpha}$ in accordance with diffusion limited cluster aggregation
by using $n_{zs} \approx  n_s$.
Now we consider the supercritical region $p > p_c$ where finite clusters and one giant cluster coexist. 
In this situation, we cannot assume that $n_{S_1}$ is an exponentially decreasing function. For simplicity, we assume that the giant cluster
is a sphere and its diameter is $S_1^{1/d_f}$. If the condition for temperature is $gz_0S_2 \ll kT \ll gz_0S_1$ for the size of the second largest cluster $S_2$,
the giant cluster performs a random walk 
at the bottom of the vessel $z\approx 0$, and finite clusters are almost uniformly distributed as shown in the Fig.~\ref{figS:schematic_boltzmann}(c).
Then, the relative collision rate $K_{S_1j}n_{S_1}n_{j}/K_{ij}n_in_j$ for $i, j \neq S_1$ would be of order 
$O(S_1^{1/d_f}/z_0)$. When $S_1^{1/d_f} \ll z_0$, the growth of the giant cluster is
successfully suppressed over extended periods. 
Specifically,
$O(N)$-sized clusters that have emerged (after some extended time interval) move to the bottom at $z\approx 0$ and 
 aggregate with the giant cluster localized there, which shall lead to a stochastic staircase, i.e.\ a type II transition. 
To check this mechanism, direct molecular dynamics simulations would be needed, which are beyond the scope of our study here. 

{\color{black}
However, a similar mechanism has been proposed recently, possibly explaining 
the emergence of early molecular life \cite{braun2013}.
The authors 
study the escalation of polymerization in a thermal gradient 
where large polymers agglomerate at the bottom of a water-filed pore \cite{braun2013}.

\subsubsection{Generalized kinetic theory kernels 
 exhibit behaviors of type III \& IV}

\begin{figure}[t!]
\includegraphics[width=0.49\linewidth]{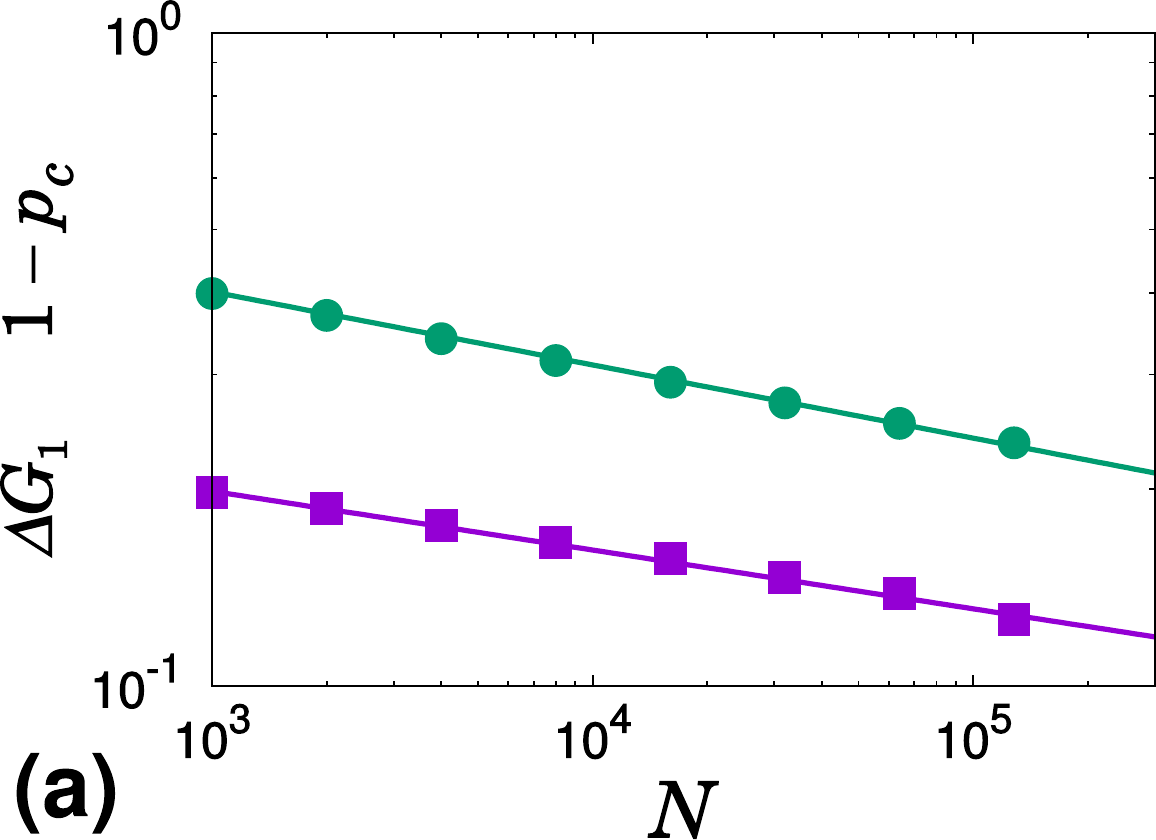}
\includegraphics[width=0.49\linewidth]{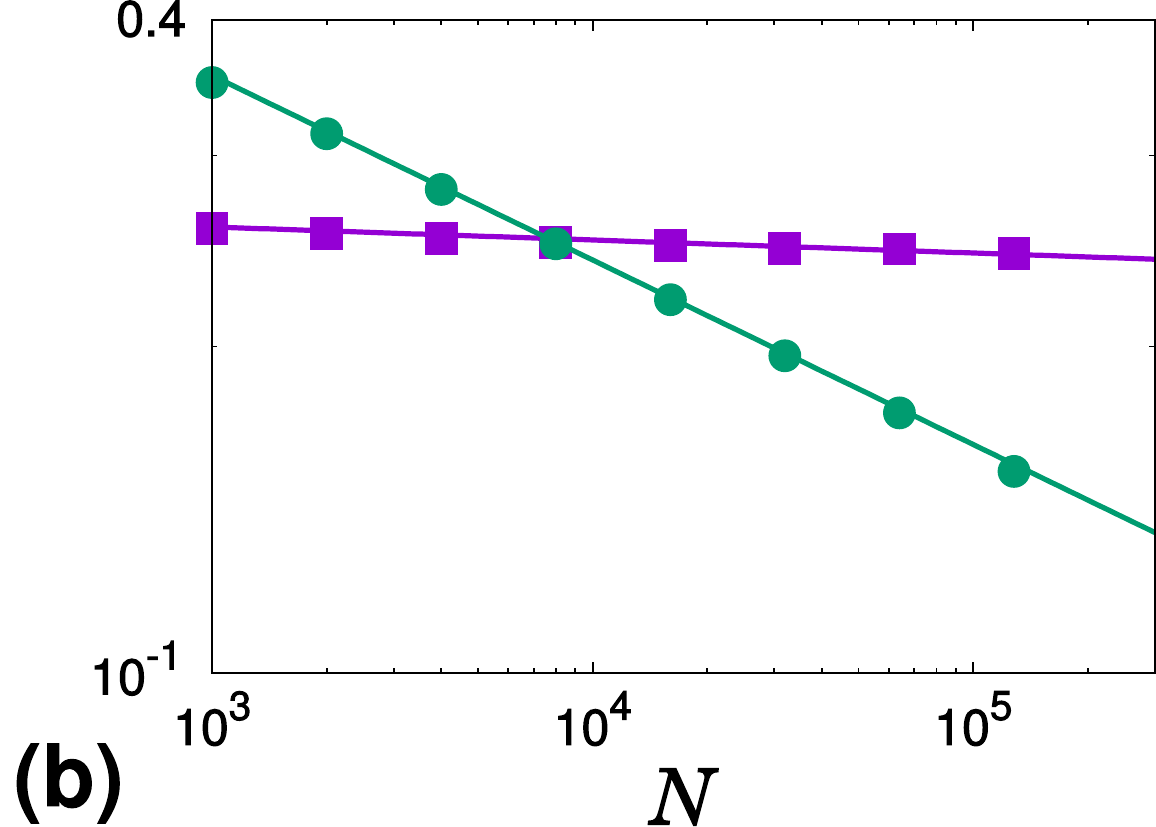}
\caption{
{\bf Type IV behavior of kernel Eq.\ (\ref{eq:general_kernel}).}
%
{\textcolor{black}{(a)}}  
Aggregation for $\alpha=1/2$, $\beta=\gamma=3/4$ and $\delta=1$ \cite{aldous1999} (no clear physical interpretation known for this kernel)
exhibits 
type IV behavior with scaling  $1-p_c(N) \sim N^{-0.11}$ ($\bullet$) and $\Delta G_1\sim N^{-0.09}$ ($\blacksquare$).
{\textcolor{black}{(b)}} System for the kernel $K_{ij}\sim (i^{1/3}+j^{1/3})^3$ ($\alpha=1/3$, $\beta=0$, $\gamma=0$ and $\delta=3$)
which describes aggregation under linear shear profile  \cite{aldous1999} exhibits
type IV behavior with scaling
$1-p_c(N) \sim N^{-0.17}$ ($\bullet$) and $\Delta G_1\sim N^{-0.012}$.
\label{SI_fig5}
}
\end{figure}

As a further
example, consider the generalized kinetic theory kernel (see, e.g.\ the review by Aldous \cite{aldous1999})
\begin{eqnarray}\label{eq:general_kernel}
K_{ij} = (i^{\alpha} + j^{\alpha})^{\delta}(ij)^{\beta}(i + j)^{-\gamma}
\end{eqnarray}
and its decomposition
\begin{eqnarray}\label{eq:general_kernel_decomposed}
K_{ij} = 
\begin{cases} 
i^{\alpha\delta+\beta-\gamma} j^{\beta} &\mbox{if }  i\gg j,\\
(i j)^{\frac{\alpha\delta}{2}+\beta-\frac{\gamma}{2}}   & \mbox{if } i\approx j,
\end{cases} 
\end{eqnarray}
with 
 exponents $\alpha$, $\beta$, $\gamma$ and $\delta$.
The exponents are specific to the particle mass $m$ but
are usually derived assuming homogenous spherical particles (in 3d) with a fixed radius $r\sim m^{1/3}$. 
However, the compactness and fractal dimension of the particle may depend on the size of the particle. 
A composite kernel can thus describe a rapid change of the fractal dimension as a function of the particle  mass.


For two choices of fixed exponents $\alpha=1/2$, $\beta=\gamma=3/4$ and $\delta=1$ (no clear physical interpretation is known for this kernel),
and $\alpha=1/3$, $\beta=0$, $\gamma=0$, and $\delta=3$ (aggregation in a linear shear profile) \cite{aldous1999}, with
\begin{eqnarray}\label{eq:shear}
K_{ij} = (i^{1/3}+j^{1/3})^3 \sim
\begin{cases} 
i^{1} j^{0} &\mbox{if }  i\gg j,\\
(i j)^{\frac{1}{2}}   & \mbox{if } i\approx j.
\end{cases} 
\end{eqnarray}

\begin{figure}[t!]
\includegraphics[width=0.6\linewidth]{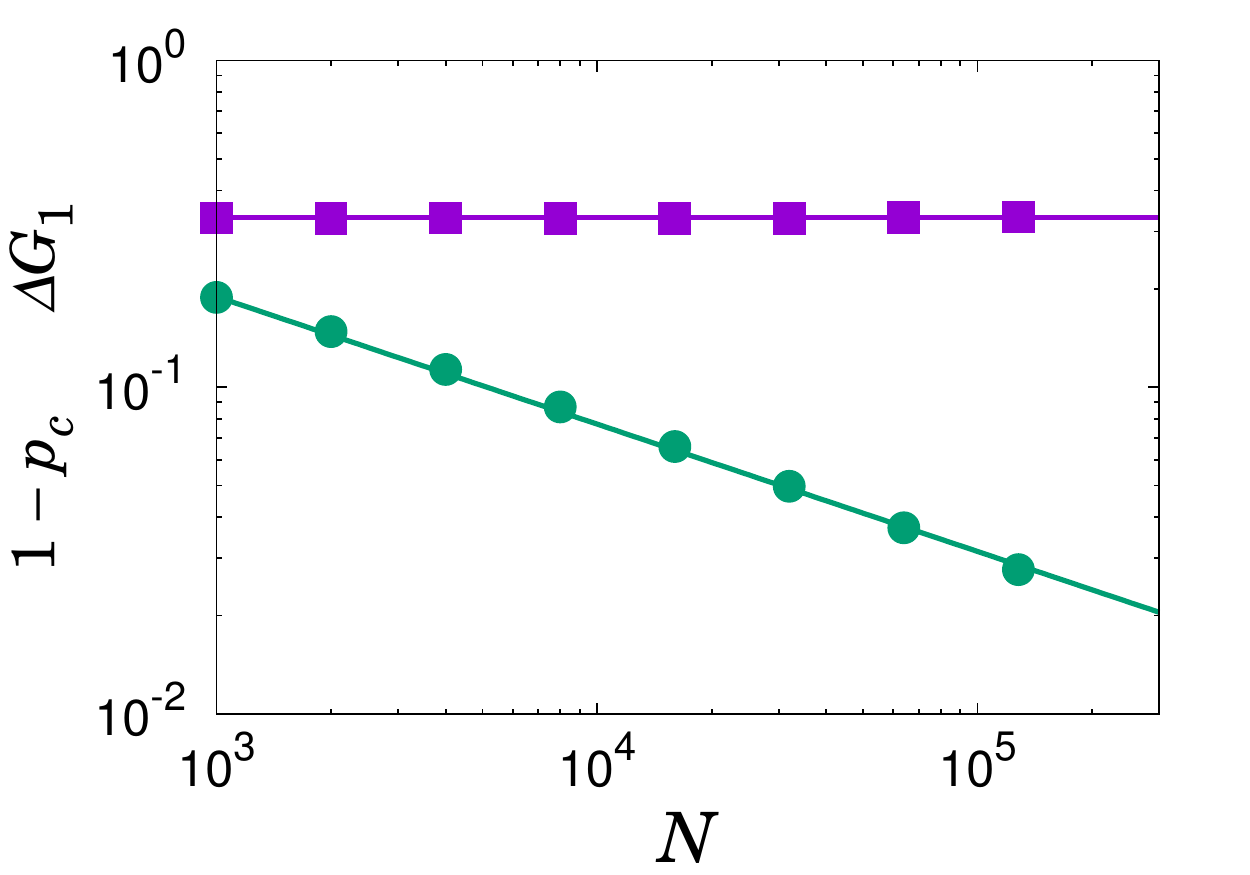}
\caption{
{\bf Type III behavior for 
cluster aggregation for a model describing aggregation in rotating systems,
using the aggregation kernel of Eq.\ (9) in \cite{saturn2015}.}
{\textcolor{black}{$1-p_c(N) \sim N^{-0.39}$ $(\bullet)$ and $\Delta G_1 \sim 0.33$ $(\blacksquare)$.}}}\label{figS:saturn}
\end{figure}

we find 
 anomalous critical behavior of type IV (see Fig.~\ref{SI_fig5}).
For $i\gg j$, the kernel in Eq.\ (\ref{eq:shear}) 
corresponds to the point $P=(0,1)$ in the phase diagram (Fig.~\ref{fig:phasediagram}).

Finally we study cluster aggregation for the kernel 
\begin{equation}\label{eq:saturn}
K_{ij}\sim (i^{1/3}+j^{1/3})^2 (i^{-1}+j^{-1})^{1/2},
\end{equation}
($\alpha=1/3$, $\beta=\gamma=-1/2$ and $\delta=2$)
which was recently suggested to describe cluster aggregation in Saturn's rings \cite{saturn2015}. 
We find scaling of the form $1-p_c \sim N^{-0.39}$  and $\Delta G_1 \approx 0.33$ (Fig.~\ref{figS:saturn}) 
suggesting type III behavior.
Limiting cases for kernels that result into mass segregation via heterogeneous time scales are easily derived,
e.g.\ 
 the {\em Saturn kernel}, Eq.\ (\ref{eq:saturn}), can be approximated by 
%
\begin{eqnarray}\label{eq:compositekernel_saturn}
K_{ij} = 
\begin{cases} 
i^{2/3} j^{-1/2} &\mbox{if }  i\gg j,\\
(i j)^{\frac{1}{12}}   & \mbox{if } i\approx j,
\end{cases} 
\end{eqnarray}
where the resulting particle size distribution based on this approximation is accurate for several orders of magnitude
for a range of system parameters, see Fig.~1 in Ref.\ \cite{saturn2015}.
A more detailed analysis is beyond our scope here, and we refer to Ref.\ \cite{saturn2015}.\\

In summary, composite kernels  
exhibit 
the full phenomenology of anomalous critical and supercritical behaviors in gelation.

\end{document}